\documentclass[fleqn,usenatbib]{mnras}

\usepackage{newtxtext,newtxmath}

\usepackage[T1]{fontenc}

\DeclareRobustCommand{\VAN}[3]{#2}
\let\VANthebibliography\thebibliography
\def\thebibliography{\DeclareRobustCommand{\VAN}[3]{##3}\VANthebibliography}

\usepackage{graphicx}	
\usepackage{amsmath}	
\usepackage{gensymb}

\usepackage{tabularx}
\usepackage{makecell}
\usepackage{enumitem}

\title[Radio AGN feedback with LOFAR]{A population-based approach to understanding radio AGN feedback with LOFAR: The LoTSS Deep Fields}

\author[J. C. S. Pierce et al.]{
J. C. S. Pierce,$^{1}$\thanks{E-mail: j.pierce3@herts.ac.uk}
F. Sweijen,$^{2}$
M. J. Hardcastle,$^{1}$
L. K. Morabito,$^{2,3}$
H. J. A. R\"{o}ttgering,$^{4}$
and R. D. Baldi$^{5}$
\\
$^{1}$Centre for Astrophysics Research, University of Hertfordshire, College Lane, Hatfield AL10 9AB, UK\\
$^{2}$Centre for Extragalactic Astronomy, Department of Physics, Durham University, Durham DH1 3LE, UK\\
$^{3}$Institute for Computational Cosmology, Department of Physics, Durham University, Durham DH1 3LE, UK\\
$^{4}$Sterrewacht Leiden, University of Leiden, 2300RA, Leiden, The Netherlands\\
$^{5}$INAF-Istituto di Radioastronomia, via Gobetti 101, 40129 Bologna, Italy\\
}

\date{Accepted XXX. Received YYY; in original form ZZZ}

\pubyear{2026}

\begin{document}
\label{firstpage}
\pagerange{\pageref{firstpage}--\pageref{lastpage}}
\maketitle

\begin{abstract}
Feedback from radio AGN jets is regularly implemented into contemporary models of galaxy evolution to offset radiative cooling in the large-scale environments in which they typically reside. While previous studies suggest that the total kinetic power output from radio AGN is sufficient for this purpose, many have relied on jet-power estimation from radio luminosities using generalised scaling relations that neglect additional information such as source size and environment. 
We here infer the cosmic evolution of radio AGN kinetic jet powers
using a physically motivated semi-analytic model for the first time.
Initial analysis on a sample of 619 radio AGN at $z < 2.5$ from LoTSS Deep Field and International LOFAR Telescope images of the Lockman Hole implies a population dominated by short-lived sources typically of lower jet power. After incorporating weighting towards shorter lifetimes in the inference models, we utilise ELAIS-N1 and Bo\"otes LoTSS Deep Field data to expand our analysis to a much larger sample of 5,187 objects, deriving jet kinetic luminosity functions and integrated kinetic luminosity densities for the radio AGN population out to $z = 2.5$. In broad agreement with previous results in the literature, we find the total power output per comoving volume to be $\sim$10$^{32}$--10$^{33}$ W\,Mpc$^{-3}$ across the full redshift range, with some suggestions of moderate positive evolution from $z$ = 0--1 and little evolution from $z$ = 1--2. These values are compatible with expectations from some cosmological models, providing strong evidence for the viability of feedback from radio AGN jets across cosmic time.

\end{abstract}




\section{Introduction}
\label{sec:intro}

Radio jets launched by active galactic nuclei (AGN) represent a key driver of the connection between galaxies and their central supermassive black holes. In the widely accepted \cite{bz77} model, jetted AGN
extract rotational energy from the intrinsic black hole spin via the strong magnetic fields in the system, which is then channelled along a relativistic collimated outflow into the host galaxy and surrounding environment. 
This energy must ultimately be dissipated, and with the amount lost to synchrotron radiation estimated to be typically two orders of magnitude lower than the mechanical power of the jet \citep{scheuer74}, much remains to be dispersed by other means.

There is strong observational evidence that AGN jets can drive multiphase outflows on galaxy scales from both detailed studies of individual sources \citep[e.g.][]{vm17,jarvis19,morganti21,murthy22,audibert23,holden23,ulivi24} and statistical studies of large samples \citep[e.g.][]{mullaney13,kukreti23,kukreti25,cr24}, with their impact on the interstellar medium also demonstrated by numerical simulations \citep[e.g.][]{muk18a,muk18b,meen22,talbot22,talbot24}. 
Beyond galaxy scales, the large energetic output supplied by radio AGN is thought to offset the radiative cooling of gas in clusters and groups, supported by the fact that radio jets have been seen to trace bubbles and cavities in the hot X-ray emitting gas in these environments \citep[see][for reviews]{mn07,mn12,gitti12,hc20}. By inhibiting the cooling of halo gas and suppressing subsequent star formation, this heating has been widely used to explain the dearth of observed high-mass galaxies relative to predictions of the galaxy luminosity function from earlier computational models, leading to the now common implementation of feedback from radio AGN in cosmological simulations of galaxy evolution \citep[e.g.][]{bow06,cro06,cro16,vog14,wein17,dave19,schaye25}.

A key element of characterising the impact of radio AGN feedback lies in determining the total energetic output of the population.
Since jet kinetic power is not directly measurable, research in this area must rely on its inference from observable quantities. One approach, as taken by \cite{will99}, is to consider the minimum energy that must be stored in the lobes of the radio source in order to produce the synchrotron radio emission observed. With assumptions on the source age, geometry and the synchrotron radiative efficiency, a theoretical relation can be formed, allowing an estimate of jet power to be obtained from observed radio luminosity. 
However, use of such a conversion requires careful consideration of the many sources of uncertainty, which principally lie in our lack of clarity on the jet particle content, the low-energy cut-off of the synchrotron-emitting electron population and the source age \citep[e.g.][]{hb14,hc20}.

Another approach is to consider the work required to inflate the bubbles and cavities observed in the hot X-ray emitting gas surrounding some radio sources. 
Since their enthalpies are derivable from spectroscopic X-ray data \citep[possibly aided by volume measurements from low-frequency radio data; e.g.][]{timm22,timm24}, these structures can act as calorimeters for the lobe plasma, which, when combined with estimates of the source age, allow intrinsic jet kinetic powers to be inferred \citep[e.g.][]{birzan04}. Comparisons with radio data show these jet power measurements to be correlated with the observed radio luminosity, which permits the construction of empirical relations that can be utilised for other radio AGN \citep[e.g.][]{birzan04,birzan08,cav10,osull11,hb14}.
However, employing these relations requires us to assume that they are universally applicable, when in reality the observed radio luminosity will be dependent on many physical factors \citep[such as source environment, distance and again particle content;][]{gs16,hard18}, with the method carrying a selection bias towards small sources in rich cluster environments \citep[][]{birzan12} and against powerful double sources with lobes bright in inverse-Compton emission \citep[e.g.][]{hc10}. In addition, the required assumptions about source age may lead to systematic errors in the kinetic powers derived in this way.
 
Another alternative for jet power inference is to compare observed radio AGN with predictions from semi-analytic dynamical models of radio source evolution. This procedure typically involves simulating large samples of forward-modelled radio sources with randomly assigned initial physical properties, including jet kinetic power, lifetime and environment, applying the constraints from the given sample selection and observations, and statistically matching mock observable properties of the resulting population with those of the observed sources \citep[cf.][]{ts15,turner18,hard18,hard19}.
This approach has several advantages: (i) the models are adaptable, so can be adjusted to consider factors such as differing jet particle content, environmental properties, observing frequencies and observational constraints; (ii) degeneracies due to age and environment are naturally accounted for; (iii) constraints from numerical simulations can be integrated to help alleviate some assumptions; and (iv) each observed source is treated individually, as opposed to a generalised relation being used. The kinetic powers inferred in this way are seen to be comparable to those obtained from cavity power relations in the higher radio luminosity regime, but to deviate significantly at lower luminosities \citep[][]{hard19}, while they agree with minimum energy estimations within a factor of a few across a broad range in radio luminosity, when appropriate normalisation is chosen \citep[][]{hard18}.

The advent of large quantities of highly sensitive radio survey data has allowed these jet power inference methods to be used to construct kinetic luminosity functions for the radio AGN population and estimate the total power output from jets. 
Several of the most recent advancements in this area have been provided by observations from the LOw Frequency ARray \citep[LOFAR;][]{vh13}. \cite{hard19} used angular size and radio luminosity measurements for a sample of 18,948 radio-loud AGN with $z < 0.8$ and $L_{144} > 10^{23}$ W Hz$^{-1}$ in Data Release 1 of the LOFAR Two-Metre Sky Survey \citep[LoTSS DR1;][]{shim19}, inferring their jet powers via comparison with the semi-analytic models of \cite{hard18} and constructing a local jet kinetic luminosity function. The integrated kinetic power output per unit comoving volume was found to exceed that of X-ray radiative losses from hot gas in groups and clusters \citep[inferred from][]{bohr14}, demonstrating that radio AGN jets are energetically capable of counterbalancing hot gas cooling, as required by cosmological models. This was supported by the cavity power relation results of \cite{igo25} based on data from the LOFAR eFEDS survey \citep[][]{igo24}, who additionally found evidence that kinetic power injection from jets provides the dominant contribution to AGN feedback relative to radiatively-driven winds in the local universe. \cite{kond23} used cavity power relations on a large sample of radio AGN detected in the LoTSS Deep Fields \citep[ELAIS-N1, the Lockman Hole, and Bo\"{o}tes;][]{tasse21,sab21}, finding only moderate kinetic luminosity evolution for the bulk of the population, but providing evidence that the total power output from the population is sufficient to offset expected losses due to radiative cooling in the redshift range $0.5 < z < 2.5$. Each of these results is supported by studies that apply cavity power or minimum energy relations to radio survey data from other telescopes \citep[e.g.][]{best14,smol17,ceraj18,but19,slaus24}. 

While these results demonstrate the feedback potential of the radio AGN population across broad periods of cosmic time, the majority of the radio sources are unresolved at the limiting angular resolution of the surveys considered, where the applicability of cavity power relations is uncertain and where poorly constrained source sizes limit the effectiveness of jet power inference from semi-analytic models.

In this paper, we use the insights gleaned from International LOFAR Telescope (ILT) data to build on the work of \cite{hard19},
by adapting and applying the \cite{hard18} semi-analytic model to a large sample of radio AGN from the LoTSS Deep Fields and investigating the jet energetics of the radio AGN population out to redshifts of $z = 2.5$. We utilise our improved measurements of the physical sizes obtained from ILT imaging of the Lockman Hole with limiting angular resolutions of 0.4 arcsec and 1.8 arcsec \citep{sweijen22,sweijen25} to improve our understanding of the radio AGN lifetime distribution, and subsequently expand our analysis to consider the much larger numbers of sources available from the 6-arcsec Lockman Hole, ELAIS-N1 and Boötes LoTSS Deep Field data. We construct jet kinetic luminosity functions for several redshift bins in the range $0.03 < z < 2.5$ and derive from these the total kinetic luminosity density of the radio AGN population at each epoch, investigating their evolution across this period of cosmic time.

The paper is organised as follows. Section~\ref{sec:data_and_sample} outlines the observational data and samples used for the project, while Section~\ref{sec:methods} describes the semi-analytic model matching approach used for the jet power inference. Section~\ref{sec:results} details the results, which are then discussed and compared with the literature in Section~\ref{sec:discussion}. The study is summarised in Section~\ref{sec:summary}. A $\Lambda$CDM cosmology with $H_0=70\ \mathrm{km}\ \mathrm{s}^{-1}\ \mathrm{Mpc}^{-1}$, $\Omega_\mathrm{m}=0.3$ and $\Omega_\Lambda=0.7$ is adopted throughout this work. We use the convention $S_{\nu} \propto \nu^{\alpha}$ for the relationship between radio flux density and observing frequency.

\section{Observational data and sample selection}
\label{sec:data_and_sample}

The samples used for this study were selected from the standard resolution ELAIS-N1, Lockman Hole and Bo\"{o}tes LoTSS Deep Field images \citep[6 arcsec;][]{sab21,tasse21} and the higher-resolution ILT images for the Lockman Hole \citep[0.4 and 1.8 arcsec;][]{sweijen22,sweijen25}, with the latter being used for our initial analysis and the former for subsequent expansion. A brief overview of the LOFAR observations and sample selection is provided in the following subsections, along with details on the source size measurements used in the analysis.

\begin{table*}
	\centering
	\caption[Summary of images used]{Summary of the LoTSS Deep Field and ILT images and sample sizes used in this work. The $N_{\rm S_{144},z}$ column indicates the numbers of objects in the fields that meet the sample selection criteria of $S_{144} > 600 \mu$Jy and $z < 2.5$, while the $N_{\rm final}$ column provides the numbers of objects in the final samples, which additionally meet the respective photometric selection criteria (see §\ref{subsec:sample_sel}). For the Lockman Hole field, the LoTSS-DF sample numbers (§\ref{subsubsec:LDF_samp}) are presented in the standard resolution image row and the Lockman Hole ILT sample numbers (§\ref{subsubsec:LH-ILT_samp}) are presented in both of the subsequent rows. Image references: 1) \cite{tasse21}; 2) \cite{sab21}; 3) \cite{sweijen22}; 4) \cite{sweijen25}.}
	\label{tab:images}
	\begin{tabular}{lccccccc} 
		\hline
		Field & \makecell{Central RA, Dec\\$[$deg$]$} & \makecell{Central RMS\\$[\mu$Jy beam$^{-1}]$} & \makecell{Beam\\ $[$arcsec$]$} & \makecell{Area\\ $[$deg$^{2}]$} & $N_{\rm S_{144},z}$ & $N_{\rm final}$ & Reference \\
		\hline
        Lockman Hole -- Standard resolution & 161.75, 58.083 & 22 & 6 $\times$ 6 & 10.28 & 2,709 & 1,991 & 1  \\
        Lockman Hole -- Intermediate resolution &  & 50 & 1.8 $\times$ 0.84 & 6.6 & 847 & 619 & 4 \\
	Lockman Hole -- High resolution &  & 25 & 0.45 $\times$ 0.40 & 6.6 & 847 & 619 & 3,4  \\
        ELAIS-N1 & 242.75, 55.00 & 20 & 6 $\times$ 6 & 6.74 & 1,625 & 1,304 & 2 \\
        Bo\"{o}tes & 218.0, 34.50 & 32 & 6 $\times$ 6 & 8.63 & 2,064 & 1,895 & 1 \\
		\hline
	\end{tabular}
\end{table*}

\subsection{ILT images -- The Lockman Hole}
\label{subsec:ILT_ims}

The broad range of baselines provided by the ILT (from $\sim$$10^2$ to $10^6$ m) permits the study of radio emission from sub-arcsecond to arcminute scales.
In addition to the LoTSS Deep Field data with the standard 6 arcsec limiting angular resolution for the Dutch LOFAR stations (see §\ref{subsec:LDFs}), we utilised the higher-resolution international-baseline ILT images of the Lockman Hole field produced by \cite{sweijen22,sweijen25} for the current work. These images, with limiting angular resolutions of 0.45 and 1.8 arcsec, were used to obtain more accurate source size measurements or size limits for objects that were unresolved in the corresponding LoTSS image for the field \citep{sweijen25}. This was crucial for constraining the semi-analytic models that produced our simulated radio sources, particularly in terms of the source lifetime distribution, which improved subsequent jet power inference for the observed sources across all of the LoTSS Deep Fields (see §\ref{sec:methods}). 

The first wide-field ILT image of the central 6.6 deg$^2$ of the Lockman Hole field was produced by \cite{sweijen22} from a single 8-hour observation, with a central resolution of 0.38 arcsec $\times$ 0.30 arcsec and central RMS sensitivity of 25 $\mu$Jy\,beam$^{-1}$. 
For the size measurement work in \cite{sweijen25}, we used these data to produce a mosaic with matching sensitivity but convolved to a restoring beam of 0.45 arcsec $\times$ 0.40 arcsec, which we hereafter refer to as the high-resolution image. Following tapering of the data, the calibration solutions from \cite{sweijen22} were also used to produce an image with an angular resolution of 1.8 arcsec $\times$ 0.8 arcsec, with central RMS noise of 50 $\mu$Jy\,beam$^{-1}$. We hereafter refer to this as the intermediate-resolution image. Further details on the ILT observations and image construction can be found in \cite{sweijen22} and \cite{sweijen25}. 
A summary of the properties of the intermediate- and high-resolution images is provided in Table~\ref{tab:images}.

\subsection{LoTSS Deep Fields -- ELAIS-N1, Bo\"otes and the Lockman Hole}
\label{subsec:LDFs}

The goal of the LOFAR Two-Metre Sky Survey (LoTSS) Deep Field observations was to provide highly sensitive low-frequency radio images for several fields with broad coverage from existing multiwavelength observations, permitting the identification and study of the host galaxies of fainter or higher redshift radio source populations over sky areas wide enough for the construction of large samples \citep[][]{kond21}. These are complementary to the shallower depth, wide-area LoTSS observations of the whole of the northern sky \citep[][]{shim17,shim19,shim22}. 

In this work we have used samples selected from the LoTSS Deep Field Data Release 1 images of the ELAIS-N1 \citep{sab21}, Lockman Hole and Bo\"otes \citep[][]{tasse21} fields. These were observed for 164 h, 112 h and 80 h respectively, with central root mean square (RMS) depths of 20, 22 and 32 $\mu$Jy\,beam$^{-1}$ being achieved. The fields were all imaged with a common circular Gaussian restoring beam with a full width at half maximum (FWHM) of 6 arcsec across the entire fields of view. A summary of the key image properties can be found alongside those of the higher-resolution ILT observations in Table~\ref{tab:images}, with further details on the observations and image construction being available in \cite{tasse21}, \cite{sab21} and \cite{kond21}.

\subsection{Sample selection}
\label{subsec:sample_sel}

\subsubsection{Lockman Hole ILT sample}
\label{subsubsec:LH-ILT_samp}

Our initial analysis was carried out on a sample of radio AGN selected from the larger sample of 2,192 radio sources from \citet{sweijen25}, which comprises all objects in the Lockman Hole high-resolution image field with LoTSS flux densities $S_{144} > 600$ $\mu$Jy. The nature of the sources was determined using the spectral energy distribution (SED) classifications derived for objects in the LoTSS Deep Fields by \citet{best23}, with the sources being placed in the following categories\footnote{Note that the use of SEDs as the basis for these classifications means that they differ from the categorisations based on optical emission line properties often used in the literature \citep[e.g.][]{but10,bh12}, which recently include probabilistic approaches (\citeauthor{drake24} \citeyear{drake24}; \citeauthor{arn25} \citeyear{arn25}; Das et al. in prep.).}: 
\begin{itemize}[label=-]
    \item \textit{Star-forming galaxies (SFGs)} -- No evidence for radiative-mode AGN emission and no radio excess relative to the radio emission expected from star formation activity;
    \item \textit{Radio-quiet AGN (RQAGN)} -- Radiative-mode AGN which do not display a radio excess;
    \item \textit{Low-excitation radio galaxies (LERGs)} -- Jet-mode AGN which exhibit a radio excess and no evidence for radiative-mode emission;
    \item \textit{High-excitation radio galaxies (HERGs)} -- Radiative-mode AGN which also exhibit a radio excess;
    \item \textit{Unclassified} -- Sources which cannot reliably be classified into any of the previous categories.
\end{itemize}
Redshifts are available for all of these classified objects from the photometric analysis of \cite{duncan21}. Sources were assigned spectroscopic redshift values whenever these measurements were available, while a hybrid template fitting and machine learning technique applied to their ultraviolet to mid-infrared photometric data was used to derive redshift estimates in all other cases \citep[see][for further details]{duncan21}.
We restrict our sample to sources with redshifts $z < 2.5$, as the SED-based classification of radiative-mode AGN becomes significantly more incomplete above this redshift \citep[see][]{best23}. Following this redshift restriction, 725 LERGs, 122 HERGs, 119 RQAGN, 842 SFGs and 114 unclassified sources remained. For the final sample of radio AGN, we considered the combined total of 847 sources from the LERG and HERG categories -- we exclude the RQAGN class due to the uncertainty that the radio emission can definitively be attributed to jets, as opposed to radiatively driven winds or other processes \citep[e.g.][]{panessa19}.

We here note that we could in principle have additionally included objects in the sample from any SED category with brightness temperatures ($T_{\rm b}$) in the high-resolution Lockman Hole image that exceed the maximum predicted value for star formation at these wavelengths \citep[e.g. $T_{\rm b}>10^6$ K, as in][]{morabito22,morabito25}. However, we choose to select only the objects with HERG and LERG SED classifications in order to maintain consistency with the sample selection for the 6-arcsec resolution LoTSS Deep Field images (§\ref{subsubsec:LDF_samp}) and with the selection of \citet[][]{kond23} based on the same data\footnote{To test the effect of adding objects with $T_{\rm b}>10^6$ K to the Lockman Hole ILT sample alongside those with HERG and LERG SED classifications, we repeated the analysis presented in §\ref{subsubsec:LH_klfs} with these objects included and compared integrated kinetic luminosity density estimates (see §\ref{subsec:klf_evol}) for the two cases. Under the assumption that all radio emission originates from radio AGN jets, it was found that adding these objects would increase the integrated kinetic luminosity density in the range $0.03 \leq z < 0.7$ by around 10 per cent. However, it is important to note that this ignores contributions to the radio emission from star-formation related processes and therefore represents an upper limit, with the true factor likely to be smaller.}.

A final requirement for the sample came from the need for optical data for the radio AGN hosts, which was important when calculating the appropriate cosmological volumes for luminosity function construction (see §\ref{subsec:local_klf}). For objects in the Lockman Hole field, the CFHT-MegaCam $g$-band data from the SpARCS survey \citep[][]{muzzin09,wilson09} were found to be most complete. We imposed a $g$-band limit of $g < 24.9$ mag on the sample of 847 HERG and LERG sources that met the flux density and redshift criteria, below which the relative uncertainties in the $g$-band fluxes all remained less than 20 per cent, producing a final sample of 619 objects (107 HERGs, 512 LERGs).
We refer to this as the Lockman Hole ILT sample from this point forward. 

\subsubsection{LoTSS Deep Field samples}
\label{subsubsec:LDF_samp}

Following initial analysis on the Lockman Hole ILT sample, we utilised the standard resolution (6-arcsec) LoTSS Deep Field images to expand the analysis to a much larger sample of radio AGN. We used the catalogues produced by \cite{best23} for the LoTSS Deep Fields, which contained 30,296, 29,691 and 17,628 sources with SED-based classifications (i.e. not "Unclassified") for the ELAIS-N1, Lockman Hole and Bo\"{o}tes fields, respectively. As with the Lockman Hole ILT sample, we restricted these catalogues to include only sources classified as LERGs or HERGs and applied a matching $S_{144} > 600$ $\mu$Jy flux density cut and $z<2.5$ redshift constraint. This left 1,625, 2,709 and 2,064 objects in the ELAIS-N1, Lockman Hole and Bo\"{o}tes catalogues, respectively. 

Following these restrictions, we again imposed the requirement of secure host galaxy detections for the radio AGN, in the optical or near-infrared (near-IR). SpARCS $g$-band data were again used for the Lockman Hole, while $z$-band data from the Hyper-Suprime-Cam Subaru Strategic Program first data release \citep[HSC-SSP;][]{aihara18} and $J$-band data from NEWFIRM observations of the Spitzer Deep Wide-Field Survey region \citep[][]{gonz10} were found to be most complete for the ELAIS-N1 and Bo\"otes fields, respectively. As for the Lockman Hole ILT sample, magnitude limits were chosen such that the relative uncertainties in the fluxes all remained less than 20 per cent, yielding restrictions of $z<23.4$ mag and $J<22.9$ mag for the ELAIS-N1 and Bo\"otes fields. A limit of $g<24.9$ mag was used for the Lockman Hole field as before. Final samples of 1,991\footnote{Note that the final Lockman Hole sample drawn from the LoTSS Deep Field data is larger than the Lockman Hole ILT sample due to the requirement for sources in the latter to lie within the smaller field of view of the ILT.}, 1,304 and 1,895 objects were obtained for the Lockman Hole, ELAIS-N1 and Bo\"{o}tes fields, respectively.
We hereafter refer to these as the Lockman Hole LoTSS-DF, ELAIS-N1 and Bo\"{o}tes samples.

\begin{figure*}
	\includegraphics[width=0.7\linewidth]{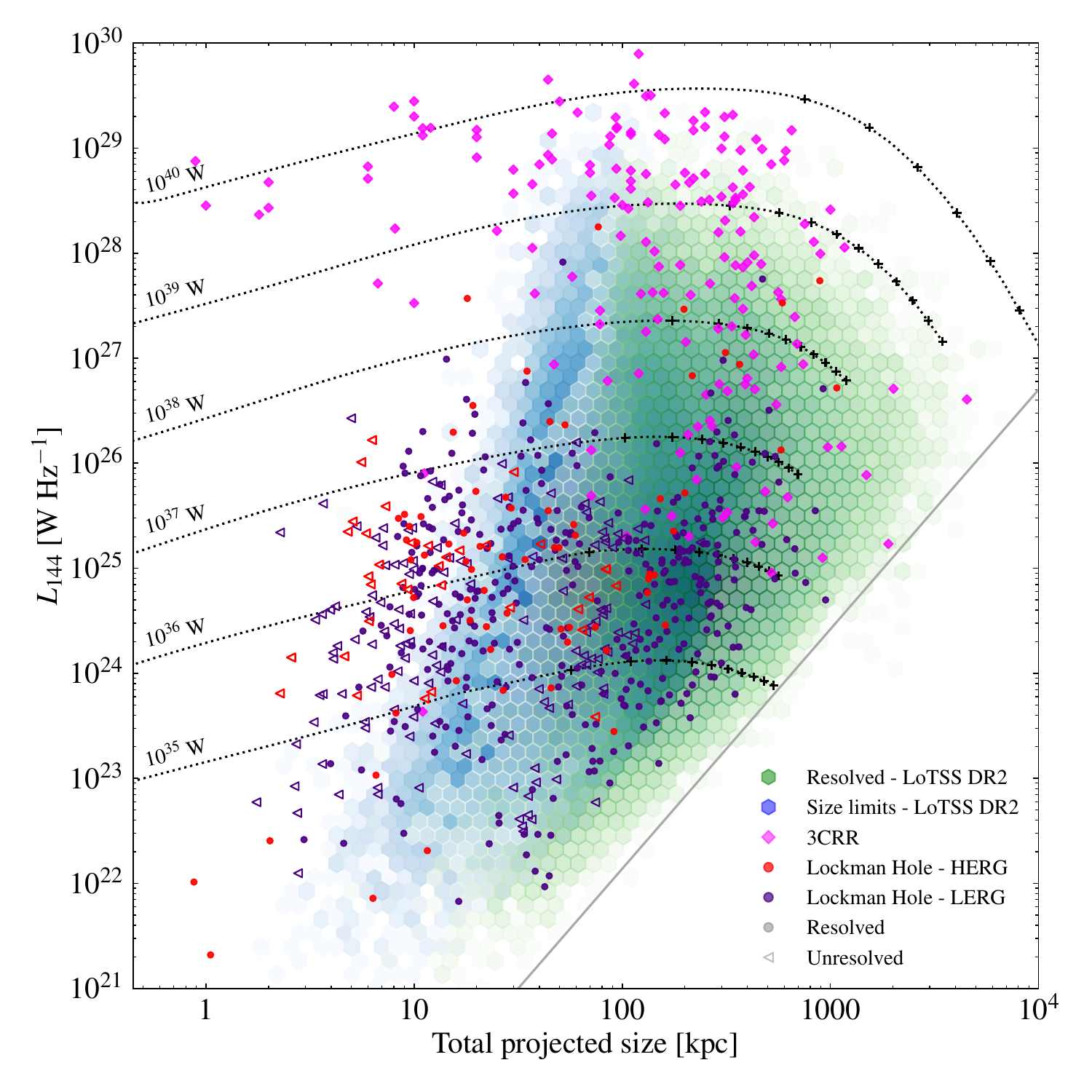}
    \caption{A power-linear size (P-D) diagram for the 619 objects in the Lockman Hole ILT sample, with the AGN subclasses from the SED classifications of \citet{best23} shown separately (HERGs in red, LERGs in indigo). The solid circles represent measured sizes, while the unfilled triangles represent size upper limits. Bins for the resolved (green) and unresolved (blue; size limits) radio AGN from LoTSS DR2 are shown, with points for the powerful radio sources in the 3CRR sample \citep{laing83} also plotted for comparison (magenta diamonds). Example evolutionary tracks from the semi-analytic models of \citet{hard18} are overplotted -- these are for $z = 0$ sources with two-sided jet powers $Q = 10^{35}, 10^{36}$... $10^{40}$ W lying in the plane of the sky and evolving through a set group environment ($M_{500} = 2.5 \times 10^{13} M_{\odot}$, $kT = 1$ keV).}
    \label{fig:pd_diagram}
\end{figure*}

\subsection{Source size measurements -- The Lockman Hole ILT sample}
\label{subsec:size_meas-LH}

Reliable measurements of the projected radio source sizes were required to improve comparisons with simulated source populations for jet power inference. The methods used to determine these sizes for the Lockman Hole ILT sample are described in detail in \cite{sweijen25}, but a brief overview is provided here.

\subsubsection{Resolved sources -- Flood-fill fitting}

The sizes of sources that were resolved in any of the LOFAR images (standard resolution, intermediate resolution or high resolution) were measured using a flood-fill fitting algorithm based on the LoMorph\footnote{\url{https://github.com/bmingo/LoMorph/}} code described by \cite{mingo19}. Image pixels with flux densities exceeding a threshold set by the highest value of $5\sigma_{\rm rms}$ or $I_{\rm peak}/50$ were included in the flood-fill mask, where $\sigma_{\rm rms}$ is the root-mean-square noise of the image and $I_{\rm peak}$ is its brightest pixel intensity. Projected physical sizes were then derived from the maximum Euclidean angular distance between any included pixels once no further pixels surrounding the defined region met the threshold. This method was used to provide measurements for 232 objects in the Lockman Hole ILT sample; 169 from the LoTSS Deep Field image, 34 from the intermediate-resolution image, and 29 from the high-resolution image.

\subsubsection{Resolved sources -- LoTSS unresolved and ILT faint/non-detected}

A different approach was taken for sources that were unresolved in the LoTSS Deep Field image but either faint or not detected in both of the ILT images. For these sources, a lower limit on their sizes was derived from a comparison between the LoTSS flux density of the source and the surface brightness limitations of the intermediate-resolution image, while the LoTSS deconvolved major axis measurement plus three times the measurement uncertainty, $\rm 2 \times (\theta^{DC}_{maj} + 3\sigma_{maj})$, was used as an upper limit. An estimate of the source size was then randomly selected from a log-uniform distribution between these two limits, motivated by the large number of sources that remained unresolved in the high-resolution image relative to those resolved in any of the three Lockman Hole images. Objects that were detected in the high-resolution image but that had flux densities $S_{\rm ILT} < 0.7\,S_{\rm LoTSS}$ were also assigned sizes in this way \citep[this factor is estimated from consideration of the typical combined flux density scale and fitting uncertainties;][]{sweijen25}. This method was used to obtain size measurements for 209 sources in the Lockman Hole ILT sample.

\subsubsection{Unresolved sources -- 2D-Gaussian fitting}
\label{subsubsec:unres_sizes_ilt}

Size estimates for sources that were unresolved in the LoTSS Deep Field image and detected but unresolved in either of the ILT images were determined by 2D-Gaussian fitting, performed using the \texttt{imfit} package in CASA \citep[][]{casa22}. The high-resolution image measurement was used for sources that were unresolved and securely detected in both ILT images with flux densities $S_{\rm ILT} > 0.7\,S_{\rm LoTSS}$, while the intermediate-resolution image measurement was used when the source was not securely detected in the high-resolution image. For the matching with simulated objects from the semi-analytic models, we used a size upper limit of $\rm 2 \times (\theta^{DC}_{maj} + 3\sigma_{maj})$\footnote{This is done to make the measurements comparable with those from the flood-fill method, with a uniform-brightness spherical or elliptical emission structure assumed so that the FWHM is equivalent to the half-power brightness. We include the $3\sigma_{\rm maj}$ to allow for uncertainty to be represented in the upper limit.}, where $\rm \theta^{DC}_{maj}$ is the deconvolved major axis and $\sigma_{\rm maj}$ its formal error. 14 intermediate-resolution measurements and 61 high-resolution measurements were used as size limits for sources in the Lockman Hole ILT sample. 

For some LoTSS-unresolved objects in the sample, issues with the ILT image quality or with the 2D-Gaussian fitting from \texttt{imfit} meant that size measurements from the ILT images were not usable. In these cases, sizes provided by 2D-Gaussian fits to the source emission from \texttt{PYBDSF} \citep[][]{mr15} performed by \cite{tasse21} and \cite{sab21} were used. Size measurements for 103 objects were obtained in this way, with the $\rm \theta^{DC}_{maj}$ and $\sigma_{\rm maj}$ values from the fits again being used to set an upper limit of $\rm 2 \times (\theta^{DC}_{maj} + 3\sigma_{maj})$ on the source size for use in later analysis.

\subsubsection{Visual inspection}

As a final review, visual inspection was used to assess the suitability of the sizes provided by the methods above and confirm the appropriate final measurements. Circles with the sizes determined by each method were overlaid on the LoTSS Deep Field and two ILT images of the objects for comparison with the maximum extents of the observed source emission \citep[see Figure 2 in][for examples]{sweijen25}. Size measurements from the intermediate-resolution image were found to be the most unreliable, due to contamination from the sidelobes of the point-spread function (PSF) and its higher RMS noise, while those from the LoTSS-DF and high-resolution images largely performed well in the relevant scenarios outlined in this subsection.

\subsubsection{P-D diagram}
\label{subsubsec:pd_diagram}

Using the LoTSS-DF luminosity densities ($L_{144}$) for the objects and the size measurements from \cite{sweijen25}, we can construct a power-linear size (P-D) diagram for the Lockman Hole ILT sample (Figure~\ref{fig:pd_diagram}), a frequently used tool for comparing populations of radio sources. The results for the HERG and LERG subsamples are overlaid on those for the AGN from LoTSS DR2 \citep{hard25}, with points for the powerful radio galaxies in the 3CRR sample \citep{laing83} and example evolutionary tracks from the models of \cite{hard18} also plotted for reference. 

We immediately see that while there are numerous resolved sources with large sizes, the sample also contains many sources of galactic or sub-galactic physical scales, with 283 objects (46 per cent) having sizes less than 30 kpc. This latter region of the diagram also contains many size limits (112; 40 per cent of objects $<$30 kpc), implying the existence of a significant population of truly small sources. A first-order comparison with the predicted evolutionary tracks from the semi-analytic models suggests that there are many objects with low kinetic jet powers in the sample. We reserve further analysis for later sections.

\subsection{Source size measurements -- The Lockman Hole LoTSS, ELAIS-N1 and Bo\"otes samples}
\label{subsec:size_meas-DF}

As for some of the LoTSS-unresolved objects in the Lockman Hole ILT sample (§\ref{subsubsec:unres_sizes_ilt}), the \texttt{PYBDSF} \citep[][]{mr15} source size measurements from \cite{tasse21} and \cite{sab21} were used for objects in the ELAIS-N1, Lockman Hole LoTSS and Bo\"otes samples, both for resolved and unresolved sources. 
For unresolved sources, upper limits on the source sizes were set at $\rm 2 \times (\theta^{DC}_{maj} + 3\sigma_{maj})$, as described in §\ref{subsubsec:unres_sizes_ilt}. For resolved sources, sizes are taken as the largest of the axes determined from moment analyses on the multiple Gaussians fitted to the source components by \texttt{PYBDSF}. Resolved and unresolved sources were separated using the equation for the envelope defined by \cite{shim19} in LoTSS Data Release 1 \citep[as also used by][]{hard19}: $S_{\rm int}/S_{\rm peak} = 1.25 + 3.1\left(\frac{S_{\rm peak}}{\rm RMS}\right)^{-0.53}$, where $S_{\rm peak}$ and $S_{\rm int}$ represent the peak and total flux densities and objects with $S_{\rm int}/S_{\rm peak}$ values above this limit were classed as resolved.
\vspace{1.5cm}

\begin{figure*}
    \includegraphics[width=\linewidth]{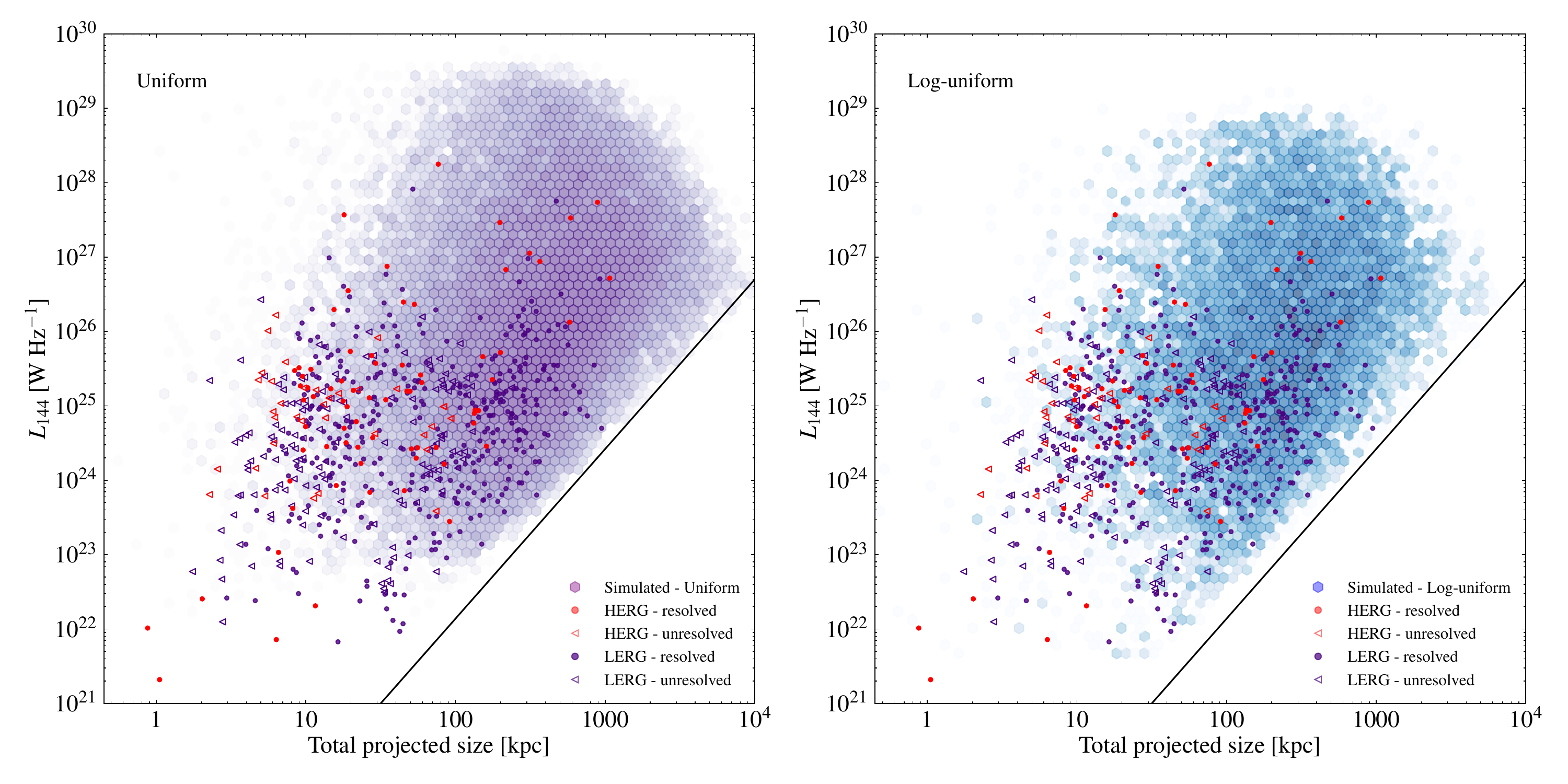}
    \caption{Comparison of realisations of the simulated source populations drawn from uniform (left; 46,897 sources) and log-uniform (right; 11,957 sources) lifetime distributions, with observational data from the Lockman Hole ILT sample overplotted. The HERG (red) and LERG (blue) subclasses from the classifications of \citet{best23} are again shown separately, with resolved (solid circles) and unresolved (unfilled triangles) sources also indicated.}
    \label{fig:lifetime_comp}
\end{figure*}

\section{Simulations and source matching}
\label{sec:methods}

The main goal of this work is to infer the underlying jet properties of the observed radio AGN population. To do this, we followed the approach of \citet{hard19} in using semi-analytic models to simulate the evolution of a large number of radio AGN with known initial conditions, then matching their mock observable properties to those of our observed sources. Details on the modelling and subsequent matching are provided in the following subsections.

\subsection{Simulated radio source populations}
\label{subsection:sim_pop}

Following \cite{hard19}, we used the semi-analytic model of \cite{hard18} to simulate radio AGN populations. 
Initialisation of the model required the random selection of several key physical properties for the radio sources. We simulated 10,000 radio sources per redshift interval of 0.1 in the range [0.05,  2.45]. Improving on previous iterations, a redshift-dependent halo mass function was employed for the random selection of the source environments \citep[from][]{tink08}, implemented in Python using the \texttt{COLOSSUS} package \citep[][]{diemer18}. The lower mass limit for the selection was also redshift dependent, set to $10^{13}$ M$_{\odot}$ at $z=0.05$ and to the equivalent values at all higher redshifts, as determined by the Tinker et al. evolution. The upper mass limit was fixed at $10^{15}$ M$_{\odot}$. Jet powers were randomly selected from a log-uniform distribution between the limits $Q= 10^{33} - 10^{39.5}$ W. The simulations were run for a total of 1200 Myr from before the cosmic epoch defined by each redshift, with random uniform radio source birth times within this interval being assigned. Source orientations relative to the line of sight were drawn under the assumption of isotropy.

The results of \cite{hard19} suggested that uniform and log-uniform source lifetime distributions could be invoked to explain the size distributions for different components of the radio AGN population, but that neither individually results in a size distribution that aptly matches that of the population as a whole. Consequently, we trialled sets of simulations considering both types of lifetime distribution and compared their mock observable properties (projected physical size and radio luminosity) to those of the Lockman Hole ILT sample, for which we have the most constraining size measurements or limits. Source lifetimes were randomly selected from the range 0.1 to 1000 Myr in each case.
Figure~\ref{fig:lifetime_comp} demonstrates how the simulated sources populate P-D space when using these two distributions, with the observational results from the Lockman Hole ILT sample overplotted for comparison. 

Observed sources were matched to the simulated sources in terms of projected physical size, 144 MHz radio luminosity and redshift, to investigate the success of the uniform and log-uniform lifetime realisations (the criteria are detailed in §\ref{subsec:obs_sim_match}).
While comparable numbers of matches are obtained when using each implementation
(both with success rates of $\sim$98 per cent),
it is found that a log-uniform lifetime distribution better populates the small physical size, low radio luminosity region of the P-D diagram where many of the observational data points lie, especially pertinent considering that many of the points in this region are size limits. It is found that 6 per cent of the simulated sources from the log-uniform realisation have sizes $<$50 kpc and luminosities less than the median of the observed sources (log$_{10}(L_{\rm 144}/\rm W\,Hz^{-1})=24.7$), compared to just 2 per cent in the uniform lifetime case, with matching success rates of 98 per cent and 95 per cent achieved in this region, respectively. The log-uniform realisation also appears to provide a better representation of the P-D space occupied by the observed sources as a whole, with an overall rate of 0.79 matches per simulated source being achieved compared to a rate of 0.59 for the uniform realisation.
As a result, we adopted a log-uniform lifetime distribution when generating simulated source populations for the current work.

\subsection{Matching to observed sources}
\label{subsec:obs_sim_match}

After the log-uniform lifetime distribution was chosen and simulated source populations were produced, final matching of simulated to observed sources was carried out to achieve the main goal of inferring the kinetic jet powers. 

As a first step, the observational constraints for the different images were taken into account. The simulated sources were filtered to match the $>$600 $\rm \mu$Jy flux density constraint imposed for all of the observed samples. The central RMS values for the three LoTSS Deep Fields were used to calculate surface brightness limits for source detection, with the flux densities and projected angular sizes for the simulated sources then used to restrict the pool to those with sufficient surface brightness to be detected. 

After applying the observational constraints, the remaining simulated objects were considered as matches to the observed sources if the following criteria were met: i) they were closer than 0.1 in redshift ($|z_{\rm obs} - z_{\rm sim}| < 0.1$); ii) the radio luminosities were consistent within 0.18 dex\footnote{This value was chosen for the matching to ensure that the observed and simulated sources were consistent within a factor of 1.5. This was found to provide the best compromise between obtaining a high number of matches and ensuring reasonable accuracy for the jet power inference.} ($\rm |log_{10}(L_{\rm 144, obs}/L_{\rm 144, sim})| < 0.18$); and iii) the projected physical sizes were consistent within 0.18 dex ($\rm |log_{10}(\theta_{\rm obs}/\theta_{\rm sim})| < 0.18$). If an observed source was deemed unresolved, constraint (iii) was adjusted, such that all simulated sources with projected physical size less than twice the deconvolved major axis plus three times the measurement uncertainty ($\rm \theta_{\rm sim} < 2 \times (\theta^{DC}_{maj} + 3\sigma_{maj})$) were considered for matching.  
Resolved and unresolved sources in the LoTSS Deep Field samples were identified as detailed in §\ref{subsec:size_meas-DF}. For the Lockman Hole ILT sample, resolved and unresolved sources were separated as described in §\ref{subsec:size_meas-LH}, with the additional stipulation that objects with projected angular sizes above 15 arcsec were deemed to be resolved. 

Once appropriate matches were identified, each observed source was assigned the Gaussian-weighted mean kinetic jet power of its matched simulated sources, with weights determined by the observed-to-simulated source separations in P–D parameter space. These values were used as the inferred jet powers in all subsequent analysis. Matches were found for 605 of the 619 objects in the Lockman Hole ILT sample (98 per cent), while 1,913 (96 per cent), 1,243 (95 per cent) and 1,687 (89 per cent) of objects in the Lockman Hole-DF, ELAIS-N1 and Bo\"otes samples were found to have matches, respectively. Only these matched sources were used for the analysis presented in §\ref{sec:results}.

\section{Results}
\label{sec:results}

\begin{figure*}
\includegraphics[width=0.45\linewidth]{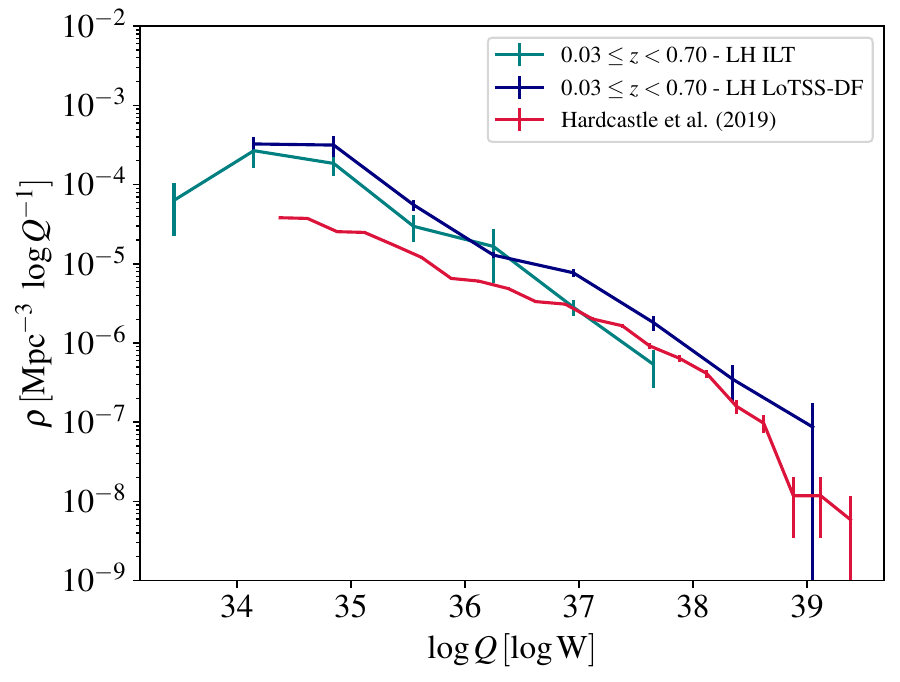}
\hspace{0.2cm}
\includegraphics[width=0.45\linewidth]{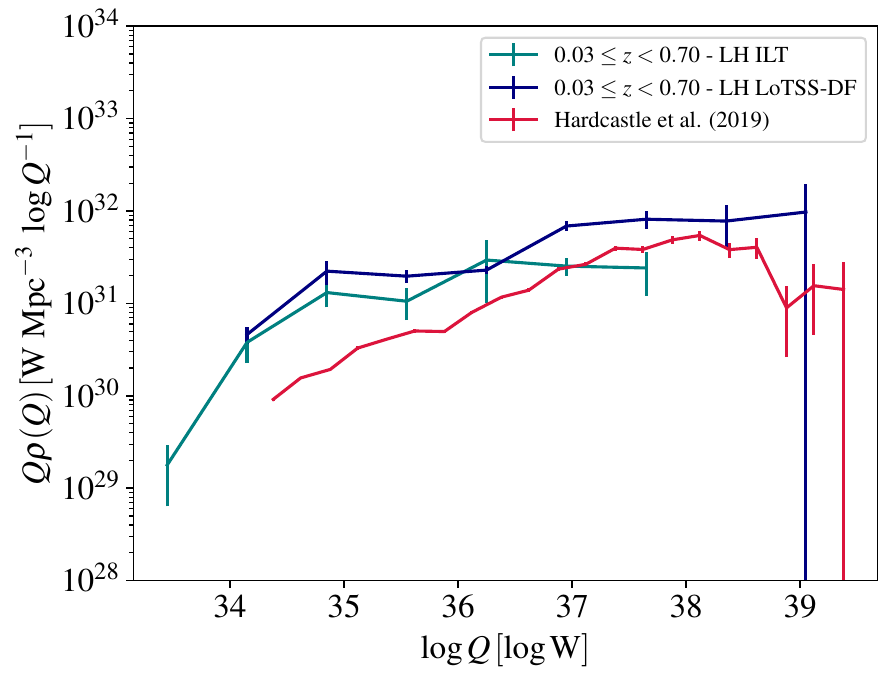}
    \caption{Kinetic luminosity functions for RLAGN with $0.03 \leq z < 0.7$, derived using the ILT (teal) and LoTSS Deep Field (blue) samples for the Lockman Hole. The results of \citet{hard19} are also shown, for comparison. Left panel: the unscaled luminosity function. Right panel: the luminosity function multiplied by $Q$, emphasising the contribution to the overall integrated luminosity density as a function of $Q$.}
    \label{fig:ilt_6arc_lh_klf}
    \label{fig:ilt_6arc_lh_qklf}
\end{figure*}

\begin{figure*}
\includegraphics[width=0.45\linewidth]{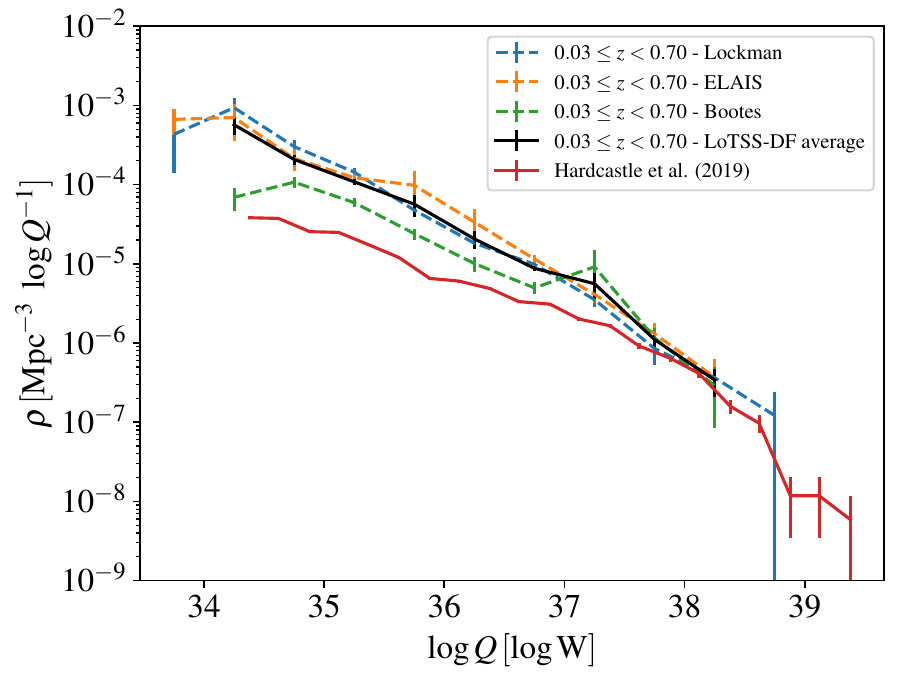}
\hspace{0.4cm}
\includegraphics[width=0.45\linewidth]{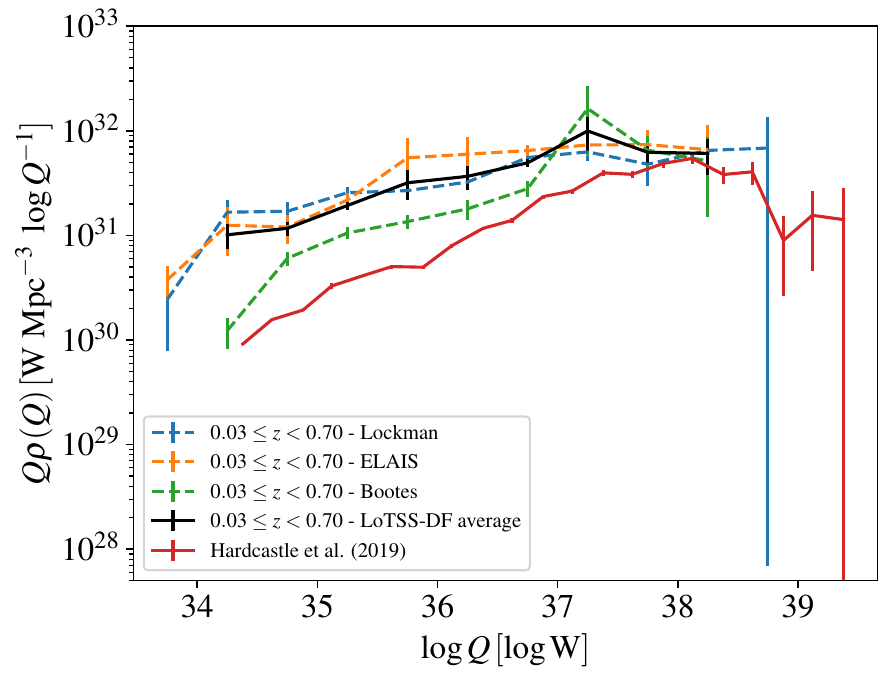}

    \caption{Kinetic luminosity functions for RLAGN with $0.03 \leq z < 0.7$, derived using the Lockman Hole (blue), ELAIS-N1 (orange) and Bootes (green) LoTSS Deep Field images. The average luminosity function for the three fields is shown in black. The results of \citet{hard19} are also shown, for comparison. Left panel: the unscaled luminosity function. Right panel: the luminosity function multiplied by $Q$, emphasising the contribution to the overall integrated luminosity as a function of $Q$.}
    \label{fig:all_df_klf}
        \label{fig:all_df_qklf}
\end{figure*}

\subsection{The local kinetic luminosity function}
\label{subsec:local_klf}

Using the inferred jet powers and known optical magnitudes and radio flux densities for the observed sources, kinetic luminosity functions could be constructed. In line with many previous studies in this area \citep[e.g.][]{williams18,kond22,hard25}, we followed the $1/V_{\rm max}$ approach to derive the luminosity functions \citep[][]{schmidt68,condon89}. The radio source number densities in each kinetic jet power bin ($\rho$) are derived using the expression $\rho = \sum_{i} 1/V_{i}$, where $V_{i} = V_{\rm max} - V_{\rm min}$ and $V_{\rm min}$ and $V_{\rm max}$ denote the volumes corresponding to the minimum and maximum distances within which a radio source could be detected. $V_{\rm min}$ was set by the lower redshift limit for the luminosity function being derived, while $V_{\rm max}$ was dependent on comparison between the optical or near-IR magnitudes and radio flux densities of the sources with the corresponding observational limits\footnote{For the optical or near-IR data, it was important to consider the redshift dependence of the part of the host galaxy SED being observed in the corresponding bandpass. As in our previous work \citep[][]{hard25}, we used the \texttt{kcorrect} code \citep[][]{br07} to obtain rest-frame SED templates and calculate the expected observed fluxes in the given bandpass as a function of redshift, comparing these values with the relevant magnitude limits in §\ref{subsec:sample_sel} to calculate $V_{\rm max}$ in cases limited by the optical or near-IR data.}, 
as well as the upper redshift limit for the luminosity function being constructed. The initial results obtained for the Lockman Hole ILT and LoTSS-DF samples are detailed in §\ref{subsubsec:LH_klfs}, with those derived from combining all three LoTSS Deep Field samples described in §\ref{subsubsec:all_DF_klfs}.

\begin{figure*}
\includegraphics[width=0.45\linewidth]{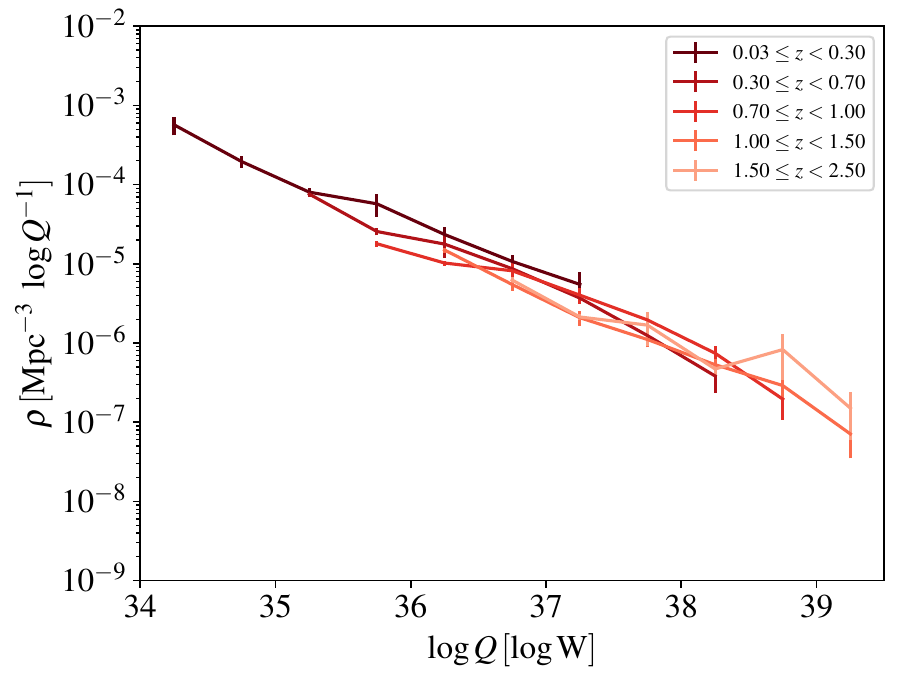}
\hspace{0.4cm}
\includegraphics[width=0.45\linewidth]{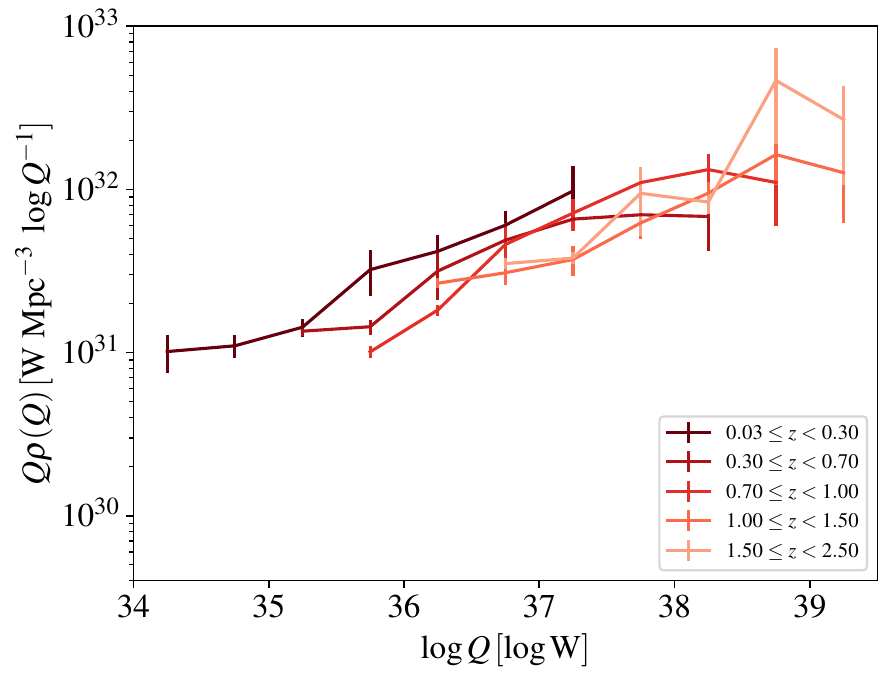}
     \caption{The evolution of the average kinetic luminosity functions for RLAGN in the redshift range $z=[0.03,2.5]$ across the Lockman Hole, ELAIS-N1 and Bootes LoTSS Deep Field images. Left panel: the unscaled luminosity function. Right panel: the luminosity function multiplied by $Q$, emphasising the contribution to the overall integrated luminosity as a function of $Q$.}
    \label{fig:all_df_klf-z_evol}
    \label{fig:all_df_qklf-z_evol}
\end{figure*}

\subsubsection{The Lockman Hole}
\label{subsubsec:LH_klfs}

We first compare the results obtained for the Lockman Hole ILT and LoTSS-DF samples. Due to the smaller size of the Lockman Hole ILT sample, we limit this comparison to kinetic luminosity functions in the local universe where both samples have suitable source counts. This also permits comparison with the results obtained by \cite{hard19} for local LoTSS DR1 AGN, which were obtained using the same jet power inference approach. 

Jet kinetic luminosity functions for the Lockman Hole ILT and LoTSS-DF samples in the redshift range $0.03 \leq z < 0.7$ are shown in Figure~\ref{fig:ilt_6arc_lh_klf} (left panel), alongside the results of \cite{hard19}. It is seen that although the luminosity function for the Lockman Hole ILT sample extends to lower jet powers when compared to that of the LoTSS-DF sample, while the latter extends to relatively higher jet powers, the luminosity functions are broadly consistent in the jet power range in which they overlap. This concordance is also seen in the $Q\rho(Q)$ plot shown in Figure~\ref{fig:ilt_6arc_lh_qklf} (right panel), which offers an impression of the total power output provided by jets of different powers based on these data.

The higher jet power extension for the LoTSS-DF sample can partly be explained by the larger sky area covered by the LoTSS Deep Field data relative to that of the ILT images (10.28 deg$^2$ and 6.6 deg$^2$, respectively), leading to the inclusion of more sources with higher radio luminosities and thus typically higher inferred kinetic powers. The additional lower jet power sources identified from the ILT sample data can be attributed to the smaller measured sizes or size limits, given that while the source lifetime is the main determinant of the source size in the semi-analytic models, smaller radio sources are still found to more frequently be of lower jet power. The broad consistency elsewhere implies that when a log-uniform lifetime distribution is used to generate the simulated source population in the semi-analytic models, as motivated by the analysis on the Lockman Hole ILT sample (see §\ref{subsection:sim_pop}), the inferred jet powers are not strongly dependent on whether the LoTSS only or LoTSS plus ILT size measurements are used. This opens up the possibility of expanding the analysis to the other LoTSS Deep Field data, where only LoTSS-based size measurements and limits are available (as detailed in §\ref{subsubsec:all_DF_klfs}).

When compared to the results of \cite{hard19}, it is seen that while the number densities typically lie within a factor of a few of those suggested by the Lockman Hole luminosity functions at higher jet powers, the Lockman Hole data suggest that sources with lower jet powers are more common (Figure~\ref{fig:ilt_6arc_lh_klf}; left panel), and hence that they provide a larger contribution to the local radio AGN power output (Figure~\ref{fig:ilt_6arc_lh_qklf}; right panel). This discrepancy can potentially be explained by the differing levels of completeness that the LoTSS DR1 and LoTSS Deep Field data have at lower flux densities, with the latter observations having much improved sensitivity relative to the former (e.g. a central RMS of 22 $\mu$Jy\,beam$^{-1}$ for the Lockman Hole LoTSS Deep Field image compared to a median sensitivity of 71 $\mu$Jy\,beam$^{-1}$ for LoTSS DR1).

\begin{figure*}
\includegraphics[width=0.7\linewidth]{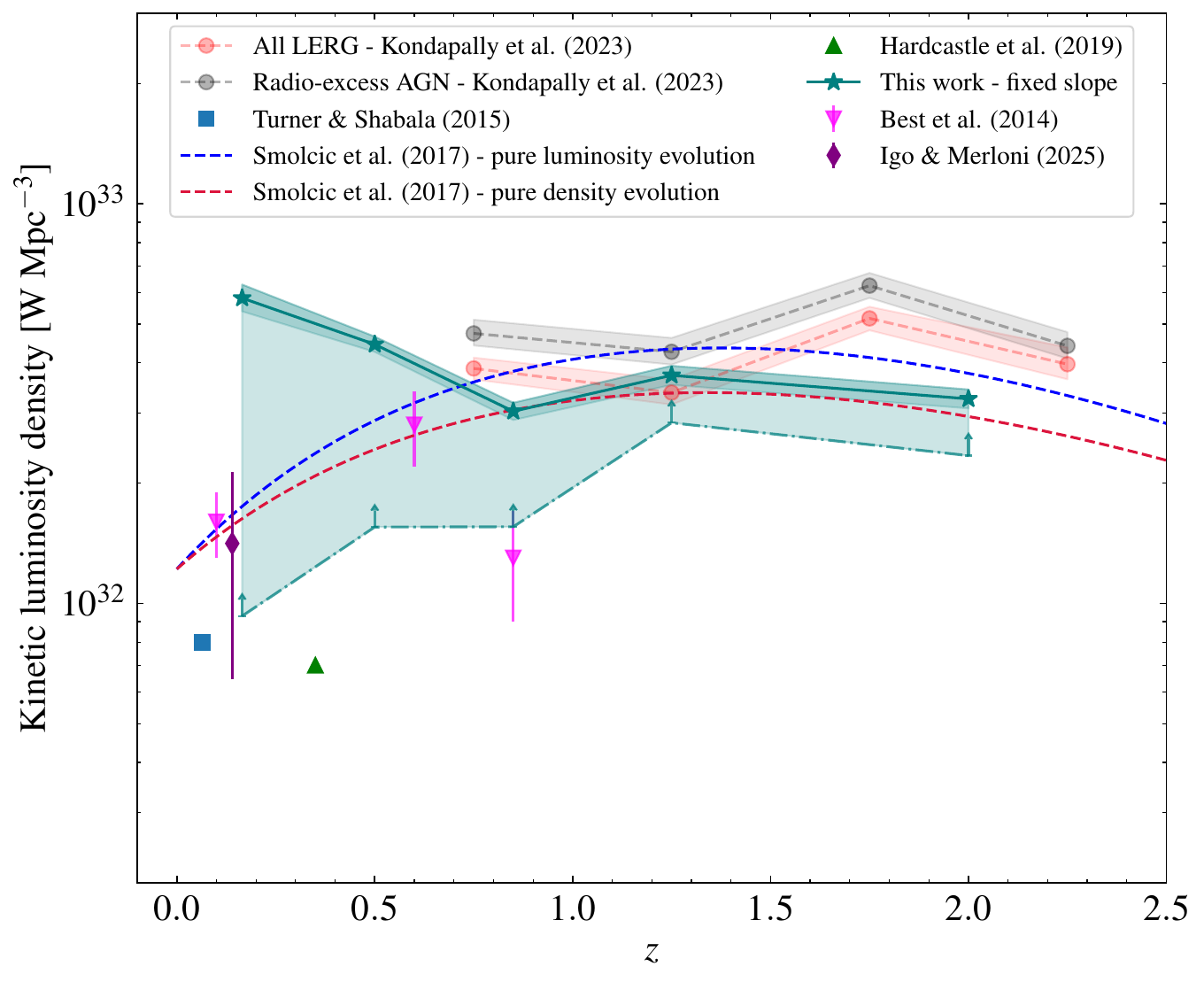}
    \caption{The integrated kinetic power density output from the radio AGN population as a function of redshift. The teal stars indicate the values derived when the kinetic luminosity functions were fitted with a slope fixed to the average value obtained from fits where both the gradient and normalisation were free parameters (see §\ref{subsec:klf_evol}, for further details on the fitting). The darker teal shading around these points indicates the 16th--84th percentile range. The teal dot-dashed line indicates the lower limit on the kinetic power density at each redshift, obtained when using integration limits given by the range of kinetic powers observed for each redshift bin, as opposed to the fixed $\rm log_{10}$$(Q)$ = 34.5 and 39.5 limits considered for the teal stars (from extrapolation of the fits). Data from other observational studies in this area are shown for comparison (also presented in Table~\ref{tab:jet_kin_pow}), which are as follows: \citeauthor{best14} (\citeyear{best14}; magenta inverted triangles); \citeauthor{ts15} (\citeyear{ts15}; light blue square); \citet{smol17} best fitting curves based on pure luminosity evolution (blue dashed line) and pure density evolution (red dashed line); \citeauthor{hard19} (\citeyear{hard19}; green triangle); \citet{kond23}, where red circles represent LERGs and grey circles represent all radio-excess AGN, with the shaded regions representing associated uncertainty ranges; and \citeauthor{igo25} (\citeyear{igo25}; purple diamond).}
    \label{fig:int_kin_dens}
\end{figure*}

\subsubsection{LoTSS Deep Fields}
\label{subsubsec:all_DF_klfs}

The broad consistency found between the kinetic luminosity functions of the Lockman Hole ILT and LoTSS-DF samples provided justification for extending the analysis to include the ELAIS-N1 and Bo\"otes LoTSS Deep Fields. This allowed for a much larger sample to be studied, and permitted investigation of the evolution of the jet kinetic luminosity functions (see §\ref{subsec:klf_evol}).

Figure~\ref{fig:all_df_klf} shows the jet kinetic luminosity functions constructed for each of the three LoTSS Deep Fields in the same redshift range as Figure~\ref{fig:ilt_6arc_lh_klf} ($0.03 \leq z < 0.7$). The average luminosity function across the Deep Fields is also shown, with the LoTSS DR1 AGN data from \cite{hard19} again presented for comparison. We see that the ELAIS-N1 and Bo\"otes luminosity functions follow the same broad trend as that of the Lockman Hole, with a steep increase towards lower jet powers and number densities that consistently exceed those derived from the LoTSS DR1 data across the majority of the jet power range covered. The average Deep Field line diverges strongly from the LoTSS DR1 line at the lowest jet powers, with the number density of order one magnitude lower for the latter at $Q = 10^{35}$ W.

Figure~\ref{fig:all_df_qklf} also shows the kinetic luminosity functions for the deep fields multiplied by $Q$. We see that the kinetic power output density remains relatively flat across the range of $Q$ considered, implying that, in contrast with the results obtained from LoTSS DR1, radio sources with lower jet powers provide an important contribution to the total energetic output of the population. Although the high kinetic power radio galaxies still provide the dominant contribution, given the increasing number of AGN sources with low radio luminosities being detected by modern surveys \citep[e.g.][]{shim19,shim22,morabito22,morabito25,baldi23,hale25}, this result would have strong implications for AGN feedback models.

\subsection{The evolution of the jet kinetic luminosity function}
\label{subsec:klf_evol}

We next constructed kinetic luminosity functions over the full redshift range of the three deep fields, $0<z<2.5$, binning in five broad redshift bins. As we have seen that the low-redshift Deep Field luminosity functions are in good agreement (§\ref{subsubsec:all_DF_klfs}), we simply use the average of the luminosity functions for the three fields in this analysis.

There is little dependence on redshift seen in the resulting LoTSS Deep Field luminosity functions (Fig.\ \ref{fig:all_df_klf-z_evol}). Although the luminosity functions span different luminosity ranges as a result of the flux density limit of the samples and the larger volumes available at high redshift, they are very similar in their general location in the plot and can be characterized roughly as power laws in $Q$ with similar normalization and slope. No positive evolution in the normalization is seen with redshift and there is little sign of a break in the kinetic luminosity function.

To allow a direct comparison with other work, we integrate the kinetic luminosity function over the jet power range $10^{34.5}$ to $10^{39.5}$ W (the latter being about the maximum jet power that is observed in the local Universe). To do this, we characterized the luminosity function in each of the five redshift bins as a power law; we fitted a power law to each, took the mean slope from these fits, and then refitted for normalization with the slope of the power law in each redshift bin fixed to the mean.
Integrating across the $Q\rho(Q)$ curves derived from these latter fixed-slope fits then allowed us to estimate the kinetic luminosity density (the total power output from radio AGN per comoving volume) as a function of redshift. 

Because the lower-redshift bins do not probe enough volume to include the most luminous sources, there is substantial uncertainty in the integral of the kinetic luminosity function (which could turn down sharply outside the range we observe), but we can derive an upper limit by extrapolating the integral over the full range between $10^{34.5}$ and $10^{39.5}$ W and a lower limit by integrating over only the range of kinetic powers observed in each redshift bin. In Fig.\ \ref{fig:int_kin_dens} we plot this whole range to illustrate the observational uncertainty, together with estimates of this quantity at different redshifts from earlier work. The implications of this plot are discussed in the following section.

\section{Discussion}
\label{sec:discussion}

\subsection{The cosmic evolution of radio AGN jet powers}

The main result from our analysis is a new estimate of the evolution of the radio AGN kinetic luminosity function and total power output over cosmic time out to $z \approx 2$, as shown in Fig.\ \ref{fig:int_kin_dens}. 
Our estimated kinetic luminosity densities agree well in terms of magnitude and suggested evolution with the results of \cite{kond23} derived from the same data, as well as with earlier analysis by \cite{smol17} on VLA-COSMOS data, both of which employ generalised radio luminosity to kinetic power relations. However, our models incorporate more physical effects, including the observed size distribution, the evolution of the halo mass function with redshift, and the dependence of inverse-Compton losses on redshift. We also see good consistency (within a factor of a few) between our low-redshift estimates and the low-$z$ estimates from \cite{ts15}, \cite{hard19} and \cite{igo25}, bearing in mind the large uncertainties on our low-$z$ results due to the small volume sampled, and the fact that we pick up more low-$z$ AGN in our current work than were identified by \cite{hard19}. Our results, combined with those of others, are consistent with moderate positive evolution of the luminosity density (by a factor of a few) between $z=0$ and $z=1$, and relatively flat or declining luminosity density beyond that.

As already noted by \cite{kond23}, numbers in the range around $5 \times 10^{32}$ W Mpc$^{-3}$ at $z=1$ are compatible with the expectations from some models of galaxy formation and evolution, although the \cite{kond23} data and models focused particularly on the LERG population, whereas we are trying to estimate the output of all jetted AGN (see \citeauthor{Hardcastle18b} \citeyear{Hardcastle18b} for the reasoning behind this decision). We already know from the work of \cite{hard19} that the kinetic luminosity density is more than sufficient to offset the observed local radiative cooling of groups and clusters in the local universe. Any \textit{excess} of kinetic luminosity density relative to the expectations from feedback models can be explained in terms of inefficient coupling of the jet to the material that actually needs to be heated in order to keep the hot halo hot and prevent star formation -- we know that large radio sources spend much of their time doing work on parts of the cluster environment where the cooling time is very long, and can even escape into voids \citep{Omma+Binney04,Hardcastle+Krause13,Oei+24}. Our $z\sim0$ estimate of the kinetic luminosity density is a factor of a few higher than that in our earlier work \citep{hard19} but, as discussed above, this is because a larger number of low-luminosity AGN are picked up in the current sample.

The large uncertainties on our results, particularly at low $z$, show that the LoTSS Deep Fields are not large enough to capture the full range of kinetic powers and environments in the powerful AGN population, and this gives rise to substantial uncertainty on the evolution of the kinetic luminosity function. At low $z$ we can address this using the population of AGN in the wide-area LoTSS survey, where the evolution of the radio luminosity function can be very well constrained \citep{hard25}. We will report on that analysis in a subsequent publication, but it will only give constraints out to $z\approx 1$ due to the requirement for detection in the DESI Legacy survey. Traditionally, the problem with wide-field radio surveys has always been the difficulty of obtaining optical IDs out to high redshift without expensive pointed optical followup, meaning that it has always been necessary to combine wide and deep observations to get the best results \citep[e.g.][]{slaus24}. However, the \textit{Euclid} mission \citep[][]{euclid25} is in the process of carrying out a wide-area optical survey which, together with complementary ground-based observations, will be the deepest wide-area optical survey yet, and should be capable of detecting the host galaxies of radio AGN out to $z\approx 2$. There will be around 6,000 square degrees of overlap between the final \textit{Euclid} northern-sky survey and the forthcoming LoTSS DR3, which will enable radio source evolution studies to be conducted out to `cosmic noon' with a sample size of around $10^6$ objects. This should give us the definitive answers to the questions posed by our existing work.

\begin{table}
	\centering
        {\renewcommand{\arraystretch}{1.6}}
	\caption{Comparison of the integrated jet kinetic power outputs per comoving volume from several radio AGN samples, as shown in Figure~\ref{fig:int_kin_dens}. References for the relevant literature data are included in the first column. The fixed slope extrapolation values from the current work (§\ref{subsec:klf_evol}) show the results of integrating the best MCMC line fits across a common range of $Q=10^{34.5}$ to $10^{39.5}$ W, with the uncertainties representing the results derived from the 16th and 84th percentile values for the fit parameters. The results calculated by only integrating across the range of $Q$ values covered by the kinetic luminosity functions in each redshift bin are also presented, representing lower limits on the integrated kinetic power density.}
	\label{tab:jet_kin_pow}
	\begin{tabular}{lcc} 
		\hline
		Reference & \makecell{Kinetic power density\\$[$W\,Mpc$^{-3}$$]$} & Redshift range\\
		\hline \\[-2.0ex]
        \cite{best14} & ($1.6$$\pm$$0.3) \times 10^{32}$ & $0 \leq z \leq 0.3$\\
         & ($2.8$$\pm$$0.6) \times 10^{32}$ & $0.5 \leq z \leq 0.7$\\
         & ($1.3$$\pm$$0.4) \times 10^{32}$ & $0.7 \leq z \leq 1.0$\\
        \cite{ts15} & $8 \times 10^{31}$ & $0.03 \leq z \leq 0.1$\\
        \cite{hard19} & $7 \times 10^{31}$ & $0.01 \leq z \leq 0.7$\\
		\cite{kond23} & &\\ 
        -- \textit{Radio-excess AGN} & $4.7^{+0.4}_{-0.3} \times 10^{32}$ & $0.5 \leq z \leq 1.0$ \\[0.7ex]
         & $ 4.3^{+0.4}_{-0.3} \times 10^{32}$ & $1.0 \leq z \leq 1.5$ \\[0.7ex]
         & $ 6.2^{+0.5}_{-0.4} \times 10^{32}$ & $1.5 \leq z \leq 2.0$ \\[0.7ex]
         & $ 4.4^{+0.4}_{-0.3} \times 10^{32}$ & $2.0 \leq z \leq 2.5$ \\[0.7ex]
        -- \textit{All LERGs} & $3.9^{+0.3}_{-0.2} \times 10^{32}$ & $0.5 \leq z \leq 1.0$ \\[0.7ex]
         & $ 3.4^{+0.3}_{-0.2} \times 10^{32}$ & $1.0 \leq z \leq 1.5$ \\[0.7ex]
         & $ 5.2^{+0.4}_{-0.3} \times 10^{32}$ & $1.5 \leq z \leq 2.0$ \\[0.7ex]
         & $ 4.0^{+0.4}_{-0.3} \times 10^{32}$ & $2.0 \leq z \leq 2.5$ \\[0.7ex]
        \cite{igo25} & $1.4^{+0.7}_{-0.8} \times 10^{32}$ & $0 \leq z \leq 0.285$\\[0.7ex]
        This work & & \\
        -- \textit{Fixed slope extrapolation} & $5.8^{+0.5}_{-0.4} \times 10^{32}$ & $0.03 \leq z \leq 0.3$\\[0.7ex]
        & $4.4^{+0.2}_{-0.2} \times 10^{32}$ & $0.3 \leq z \leq 0.7$\\[0.7ex]
        & $3.0^{+0.2}_{-0.1} \times 10^{32}$ & $0.7 \leq z \leq 1.0$\\[0.7ex]
        & $3.7^{+0.2}_{-0.2} \times 10^{32}$ & $1.0 \leq z \leq 1.5$\\[0.7ex]
        & $3.3^{+0.2}_{-0.2} \times 10^{32}$ & $1.5 \leq z \leq 2.5$\\[0.7ex]
        -- \textit{No extrapolation} & $9.3 \times 10^{31}$ & $0.03 \leq z \leq 0.3$\\
        & $1.6 \times 10^{32}$ & $0.3 \leq z \leq 0.7$\\
        & $1.6 \times 10^{32}$ & $0.7 \leq z \leq 1.0$\\
        & $2.8 \times 10^{32}$ & $1.0 \leq z \leq 1.5$\\
        & $2.3 \times 10^{32}$ & $1.5 \leq z \leq 2.5$\\
		\hline
	\end{tabular}
\end{table}

\subsection{Modelling radio AGN jet dynamics and jet power inference}

A key feature of our work here and earlier \citep{hard19} is that we are trying to use a physically motivated model for AGN evolution to infer jet power, rather than using simple scaling relations with radio luminosity such as those of \cite{will99} or \cite{cav10}. Discrepancies between the integrated kinetic luminosity densities in different studies can arise as a result of different choices of scaling relations, or freely adjustable normalization parameters in the scaling relations used. In principle, at least, our approach improves on these methods because it directly uses all available physical information (so far, source size and redshift as well as radio luminosity, but environmental and integrated spectral index information will be included in later implementations) and the free parameters that the model does have can be largely tied to observations\footnote{A comparison between the results of our jet power inference method and several generalised scaling relations from the literature can be found in \cite{hard19}. We have verified that the current data compare similarly, and we find no significant variation with redshift or between the jet powers inferred for our HERGs and LERGs.}.

There are, however, inevitable downsides to attempts to use a physically motivated model. One of them is that it leads us to face some of the remaining unknown physical properties of the sources (conveniently hidden in the uncertainty factors of the scaling relations): for example, our lack of environmental information for the sources in the current sample leads us to assume a particular prior for the halo masses of their environments that may not be correct, and that would translate to a systematic shift in the inferred jet powers (in the sense that if all the environments were systematically richer than in our prior, the inferred jet powers would all be systematically high relative to their true values). We will return to jet power inference making use of available environmental constraints when we consider the kinetic luminosity function of the LoTSS DR2 AGN (Pierce et al in prep.), but obtaining environmental information out to high redshifts is challenging. Another issue is the varying particle content in real radio AGN \citep{Croston+18}; while our models are calibrated to powerful FRII-type radio galaxies in which the radiating electrons and positrons appear to dominate the lobe pressure, it seems impossible to evade the conclusion that a substantial amount of the jet power in low-power objects goes into heating entrained baryonic material from stellar winds and possibly the external environment, and in that case the radio luminosity for a given jet power would substantially decrease, meaning that we would be substantially underestimating the true kinetic luminosity density contribution from low-power sources. Other physical effects, such as jet precession due to binary supermassive black holes \citep{Krause+25} are also not represented in the models.

In addition, there are multiple analytic models of radio source evolution available to do this inference \citep[e.g.][]{turner23,bp25}, and so far we have only used our own. While these latest models share several fundamental assumptions they differ in detail, with differing prescriptions for key physical aspects such as early-time versus late-time evolution, the pressure profile of the ambient medium and the predicted radio luminosities resulting in divergences in the evolutionary tracks of the radio sources through P-D space \citep[see discussions in][]{ts23,bp25}. This would lead to differences in the kinetic jet powers inferred, subsequently affecting the kinetic luminosity functions and integrated luminosity densities obtained. A detailed quantitative comparison of the different available models and an understanding of their biases with respect to each other would be very valuable; we leave this for future work.

Finally, we note that our whole approach of inferring kinetic luminosity functions is an attempt to meet cosmological models halfway by providing them with a number that is comparable to something that can be extracted relatively easily from simulations (the power in `AGN feedback', however that is implemented in the simulation). In many ways it would be better for the radio source modelling to be part of the cosmological simulations themselves, allowing full forward modelling of observables like the radio luminosity function along with the observational selection effects on the population. Efforts to identify radio AGN and produce radio luminosity functions by post-processing simulations
\cite[e.g.][]{Thomas+21} are encouraging, but are based on relatively crude assumptions about the relationship between accretion and radio luminosity. Incorporating the full range of radio AGN physics, starting from our best understanding of the jet generation mechanism and its relationship to black hole spin, is a key requirement for exploiting the constraints on the radio source population that are now being provided by sensitive surveys with LOFAR, MeerKAT and ASKAP, and in the near future with the Square Kilometer Array (SKA).

\section{Summary and conclusions}
\label{sec:summary}

Feedback from radio AGN jets is a 
feature of modern models of galaxy evolution, commonly invoked to counterbalance radiative cooling in large-scale cosmological environments and match the observed galaxy luminosity function. Although previous work in this area suggests that the radio AGN population carries enough power in jets to provide this regulation, 
much remains to be understood about how this mode of feedback operates in detail, on various physical scales and across cosmic time.
In this work, we have used large samples of radio AGN selected from the Lockman Hole, ELAIS-N1 and Bo\"otes LoTSS Deep Fields to investigate the cosmic evolution of radio AGN jet kinetic powers out to $z=2.5$. In contrast with the generalised radio luminosity scaling relation methods used in most previous studies, we infer the kinetic powers using a physically motivated semi-analytic model; the first time this evolution has been explored using this method out to higher redshifts.

Our initial analysis was carried out on a sample of 619 radio AGN selected from LoTSS Deep Field and International LOFAR Telescope (ILT) images of the Lockman Hole with limiting angular resolutions of 6 arcsec, 1.8 arcsec and 0.4 arcsec -- the Lockman Hole ILT sample. Utilising size measurements obtained for these objects in our previous work \citep[][]{sweijen25}, we found that the semi-analytic models required a lifetime distribution that was heavily weighted towards short-lived sources in order to match the large number of physically small objects in the sample. Using our selection of a log-uniform lifetime distribution, we found that the local kinetic luminosity function ($z<0.7$) derived for sources in this sample showed good agreement with that obtained when using a sample selected from the LoTSS Deep Field observations alone. Based on this finding, we expanded our analysis to consider further large samples of sources selected from the LoTSS Deep Field data for the ELAIS-N1 and Bo\"otes fields.

After updating the inference models of \cite{hard19} with the source lifetime distribution implied by the Lockman Hole ILT results, we made estimates of the kinetic luminosity function and its integral, the kinetic luminosity density, over the redshift range $z=0$ to $z=2.5$, using the full combined sample of 5,187 objects selected from the LoTSS Deep Field data. Encouragingly, broad agreement between these results of and those from the various different methods (mostly making use of simple scaling relations) that have been used by previous works, both using the LOFAR data and using other surveys. All the available data are consistent with a moderate positive evolution of the kinetic luminosity density from AGN between $z=0$ and $z=1$, followed by a period where it is roughly constant between $z=1$ and $z=2$. The kinetic luminosity values we obtain are of order 10$^{32}$ to 10$^{33}$ W\,Mpc$^{-3}$ across the full redshift range studied, consistent with the expectations of some cosmological models of galaxy formation and evolution.

In future work, we will utilise the high sensitivity provided by telescopes such as LOFAR, MeerKAT, ASKAP, and in the near future the SKA, alongside the wide-field, high-resolution images of the ILoTSS survey to conduct further comprehensive investigation of feedback from the radio AGN population across broad ranges of size and radio luminosity. The next generation of deep, wide-area optical survey data from facilities such as \textit{Euclid} will also enable the improved detection and characterisation of the radio AGN host galaxies. In conjunction with this, we will use constraints on the environments of local radio sources provided by LOFAR, among other factors, to improve our characterisation of radio source evolution using semi-analytic modelling. Comparison between the results of the existing inference models in the literature will also help in this aspect.

\section*{Acknowledgements}

JCSP and MJH acknowledge support from the UK STFC under grants [ST/V000624/1] and [ST/Y001249/1]. LKM is grateful for support from a UKRI FLF [MR/Y020405/1] and from STFC [ST/V002406/1]. FS appreciates the support of the UK STFC under grant [ST/Y004159/1]. 

LOFAR is the Low Frequency Array, designed and constructed by ASTRON. It has observing, data processing, and data storage facilities in several countries, which are owned by various parties (each with their own funding sources), and which are collectively operated by the ILT foundation under a joint scientific policy. The ILT resources have benefited from the following recent major funding sources: CNRS-INSU, Observatoire de Paris and Université d’Orléans, France; BMBF, MIWF-NRW, MPG, Germany; Science Foundation Ireland (SFI), Department of Business, Enterprise and Innovation (DBEI), Ireland; NWO, The Netherlands; The Science and Technology Facilities Council, UK; Ministry of Science and Higher Education, Poland; The Instituto Nazionale di Astrofisica (INAF), Italy.

This work made use of \texttt{Astropy} (\href{http://www.astropy.org}{http://www.astropy.org}), a community-developed core Python package and an ecosystem of tools and resources for astronomy \citep{astropy:2013, astropy:2018, astropy:2022}, of \texttt{TOPCAT} \citep[][]{topcat} and of \texttt{Matplotlib} \citep[][]{matplotlib}.

This research made use of the University of Hertfordshire high-performance computing facility and the LOFAR-UK computing facility located at the University of Hertfordshire (https://uhhpc.herts.ac.uk) and supported by STFC
[ST/P000096/1].

\section*{Data Availability}

All data used for this project are publicly available through the LOFAR Surveys Key Science Project website, \href{https://lofar-surveys.org/}{https://lofar-surveys.org/}.



\bibliographystyle{mnras}
\bibliography{refs} 

\bsp	
\label{lastpage}
\end{document}